\documentclass[conference]{IEEEtran}
\IEEEoverridecommandlockouts
\usepackage[colorlinks,citecolor=blue,urlcolor=blue,bookmarks=false,hypertexnames=true]{hyperref} 
\usepackage{cite}
\usepackage{amsmath,amssymb,amsfonts}
\usepackage[table,xcdraw]{xcolor}
\usepackage{graphicx}
\usepackage{textcomp}
\usepackage{svg}
\usepackage{float}
\usepackage{amsmath}
\usepackage{multicol}
\usepackage{array}
\usepackage[noend]{algpseudocode}
\usepackage[ruled, lined, linesnumbered, commentsnumbered, longend]{algorithm2e}
\usepackage{multirow}

\usepackage{colortbl}
\newcommand{\mc}[2]{\multicolumn{#1}{c}{#2}}

\definecolor{blue1}{HTML}{CBD4DF}
\definecolor{blue2}{HTML}{A0B4C7}
\definecolor{blue3}{HTML}{7998B3}
\definecolor{blue4}{HTML}{5181A1}
\definecolor{blue5}{HTML}{156E92}
\definecolor{blue6}{HTML}{486B84}
\definecolor{green1}{HTML}{D9EDDC}
\definecolor{green2}{HTML}{B6DDBE}
\definecolor{green3}{HTML}{93CFA1}
\definecolor{green4}{HTML}{6DC288}
\definecolor{green5}{HTML}{3CB76D}
\definecolor{green6}{HTML}{AEDBC2}
\definecolor{green7}{HTML}{C2E3D0}
\definecolor{green8}{HTML}{D5ECDE}

\definecolor{green9}{HTML}{E4F3EA}
\definecolor{greens}{HTML}{ECF6EF}
\definecolor{greens1}{HTML}{FAFDFA}

\definecolor{greens2}{HTML}{BCE2CE}
\definecolor{greens3}{HTML}{CDE8D9}
\definecolor{greens5}{HTML}{ECF6F0}
\definecolor{gold1}{HTML}{EFDFC8}
\definecolor{gold2}{HTML}{E3C69C}
\definecolor{gold3}{HTML}{D84F74}
\definecolor{gold4}{HTML}{CD994D}
\definecolor{gold5}{HTML}{C3862B}

\definecolor{golds}{HTML}{F1E4D1}
\definecolor{golds1}{HTML}{F6EDE1}

\definecolor{gold_fig}{HTML}{F8D287}
\definecolor{green_fig}{HTML}{EBF5EC}

\usepackage{lipsum}

\def\BibTeX{{\rm B\kern-.05em{\sc i\kern-.025em b}\kern-.08em
    T\kern-.1667em\lower.7ex\hbox{E}\kern-.125emX}}
    

\begin{document}


\title{A Two-Stage Efficient 3-D CNN Framework for EEG Based Emotion Recognition 
\thanks{* Equal Contributions}
}




\author{\IEEEauthorblockN{Ye Qiao*}
\IEEEauthorblockA{\textit{Department of Electrical Engineering}\\\textit{and Computer Science}\\
\textit{University of California, Irvine}\\
Irvine, California, USA\\
yeq6@uci.edu}
\and
\IEEEauthorblockN{Mohammed Alnemari*}
\IEEEauthorblockA{\textit{Department of Electrical Engineering}\\\textit{and Computer Science}\\
\textit{University of California, Irvine}\\
Irvine, California, USA\\
malnemar@uci.edu}
\and
\IEEEauthorblockN{Nader Bagherzadeh, \textit{Fellow, IEEE}}
\IEEEauthorblockA{\textit{Department of Electrical Engineering}\\\textit{and Computer Science}\\
\textit{University of California, Irvine}\\
Irvine, California, USA\\
nader@uci.edu}
}

\maketitle

\begin{abstract}
This paper proposes a novel two-stage framework for emotion recognition using EEG data that outperforms state-of-the-art models while keeping the model size small and computationally efficient. The framework consists of two stages; the first stage involves constructing efficient models named EEGNet, which is inspired by the state-of-the-art efficient architecture and employs inverted-residual blocks that contain depthwise separable convolutional layers. The EEGNet models on both valence and arousal labels achieve the average classification accuracy of 90\%, 96.6\%, and 99.5\% with only 6.4k, 14k, and 25k parameters, respectively. In terms of accuracy and storage cost, these models outperform the previous state-of-the-art result by up to 9\%. In the second stage, we binarize these models to further compress them and deploy them easily on edge devices. Binary Neural Networks (BNNs) typically degrade model accuracy. We improve the EEGNet binarized models in this paper by introducing three novel methods and achieving a 20\% improvement over the baseline binary models. The proposed binarized EEGNet models achieve accuracies of 81\%, 95\%, and 99\% with storage costs of 0.11Mbits, 0.28Mbits, and 0.46Mbits, respectively. Those models help deploy a precise human emotion recognition system on the edge environment.

\end{abstract}


\section{Introduction}
The Deep Neural Network achieves incredible results in computer vision, speech recognition, and natural language processing \cite{lecun2015deep}. Deep neural network models also perform exceptionally well for Brain-computer Interaction (BCI) and Human-computer Interaction (HCI) fields. Electroencephalogram (EEG) signals are used in a variety of BCI applications such as control prosthetics, neurofeedback, and emotion Recgnition \cite{alkawadri2019brain}.
In real-time, a BCI system records EEG signals in a non-invasive manner and produces a message or computational command from the recorded signals. A BCI system is made up of three different components: sensors (mostly electrodes mounted on the scalp to record EEG signals); translation and communication (mostly translating EEG signals to command or computational language); real-time actions (actions based on EEG signals).
Convolutional Neural Networks (CNNs) produce promising results in computer vision applications such as image recognition, object detection, and semantic segmentation. There are various types of CNNs, but in terms of dimensionality, 1-D CNNs are the most commonly used for time series applications such as human activity identification \cite{srinivasamurthy2018understanding} and physiological signals as shown in \cite{kiranyaz20211d}. 2-D CNNs are the most commonly used for image data and computer vision applications such as image classification and segmentation. 3-D CNNs, which are mostly useful for volumetric data, have been adopted successfully in video analysis and object recognition tasks \cite{ji20123d, karpathy2014large}. 3-D CNNs take 3-dimensional inputs and apply 3-dimensional filters to them. The filters will move along three axes to form 3D shape outputs. Long sequence data such as video, audio, electrocardiogram (ECG), and electroencephalogram (EEG) signals can benefit from the extra convolutional dimension due to the spatiotemporal correlations between data segments in the long sequence.

Emotions play an important role in human reasoning, and are linked to rational decision making, perception, human interaction, and even human intelligence itself \cite{suhaimi2020eeg}. There are various methods existing for modeling human emotions, and one of the most effective approaches is using multiple dimensions or scales for emotion categorization. In such a model, emotions are defined by two major perception dimensions: valence and arousal. Arousal ranges from low to high, whereas valence ranges from positive to negative. For example, fear has a negative valence and high arousal, whereas excitement has a positive valence and high arousal.
Electroencephalography (EEG) signals are brain waves that measure eclectic field behaviors from the human scalp. They naturally can be applied to human emotion recognition due to their reflection of human response and linkage to the cortical activities \cite{salama2018eeg}. 
Several works of literature studied EEG-based emotion recognition by extracting EEG features using deep neural network algorithms. These methods demonstrated high recognition accuracies compared to classical machine learning models but still required feature extraction and selection prior to the classifier \cite{zheng2015revealing} \cite{alnemari2017integration}.
Using CNN for emotion recognition is not novel. A 2-D CNN used for emotion recognition of power spectrum density features (PSD) from the original EEG signals as input to the neural network, demonstrated good results \cite{qiao2017novel} and \cite{moon2018convolutional}. However, CNN typically requires a large number of computational resources, and computing power spectrum density is not efficient. As a result, deploying these types of models on low resource constrain devices is not possible. In this paper, we proposed a novel two-stage 3-D convolutional neural networks framework for emotion recognition without special signal transformations on EEG data. Our method significantly outperformed the state-of-the-art result while keeping the model size small and computationally efficient. Thus allowing model deployment on the edge environments where devices have limited resources.

\section{Related Works}
A variety of approaches, including traditional machine learning and deep learning algorithms, were introduced to identify and classify emotions. Traditional machine learning algorithms require feature extraction and selection before applying the classifier. The Support Vector Machine (SVM) has been used on the publicly available dataset DEAP to identify the valence and arousal perceptions using feature vectors based on statistical measurements of the frequency bands in the EEG signal, and achieved 67\% by using all features \cite{samara2016feature}. A deep belief network with glia chain (DBN-GC) was also adopted in this area. They extracted the intermediate representation of raw EEG signals from each domain separately, and then used the glia chain to mine correlation information. Lastly, they fused all information together using the Restricted Boltzmann Machine (RBM) to implement emotion recognition that achieved 75.92\% and 76.83\% on arousal and valence, respectively \cite{chao2018recognition}. On the other hand, the deep learning method proposed by \cite{gunes2013categorical} with a 2-D CNN achieved comparable performance as the SVM. However, there are two significant drawbacks of applying 2-D CNNs to raw EEG signals: covariance shift and unreliability of emotional ground truth. The covariance shift is the difference in statistical distribution between training and testing data, which is severe in EEG signals due to its non-stationary nature of the signal \cite{jirayucharoensak2014eeg}.

Usually, raw EEG signals are segmented into several input sequences to augment the data. Emotion EEG trials should correspond to their ground truths, which are self-reported. The difference between the average of the segmented signals and ground truths causes the unreliability of each epoch which influences the model training \cite{gunes2013categorical}. To address those two issues, the 3-D CNN structure was introduced because of its ability to simultaneously extract spatial and temporal features. Salama et al. \cite{salama2018eeg} proposed a 3-D CNN strategy for EEG-based emotion recognition with a data augmentation method by adding normalized random Gaussian noise and achieved a better result compared to the previous methods. Yang et al. \cite{yang2019multi} developed a different multi-column CNN structure whose prediction is produced by a weighted sum of the decisions from all individual recognizing modules and obtained around 90\% on both valence and arousal labels. Zhao et al. \cite{zhao20203d} constructed another 3-D CNN model with reshaping channel matrices that achieved the state-of-the-art result at 96.43\% and 96.61\% on valence and arousal respectively. However, the large parameters counts of the redundant model prevented its usage in practice.

\section{EEG signal}

EEG signals can be classified into different ranges based on their frequency. Delta waves are the slowest and most common in deep sleep and relaxation. Theta wave is commonly associated with Rapid Eye Movement (REM) sleep, as well as learning and memory retention. Alpha wave is linked to sensory-motor and memory function. The most common wave is the beta wave, which is associated with alertness and awareness as well as day-to-day problem-solving tasks. Gamma is the fastest wave and is associated with alertness and high virtues such as altruism and compassion \cite{ramadan2015basics}.

\subsection{Emotions Recognition}
Emotions vary from one person to another. Several methods are presented in the literature for modeling human emotions. One model depicts simple emotions such as happiness and sadness \cite{weiner1986attribution}, while the other depicts fear, anger, depression, and satisfaction \cite{izard1984emotions}. Another modeling method employs several measurements or scales to categorize human emotions \cite{posner2005circumplex}. The most prevalent human emotion model uses two key dimensions to model the emotions which are valence and arousal. Valence varies from positive to negative, while arousal ranges from low to high. Fear, for example, has negative valence and high arousal, while excitement has positive valence and high arousal \cite{reuderink2013valence}.

In the frequency domain, power features are often used in researches. The alpha band power spectral density (PSD) of EEG correlates with valence \cite{verma2014multimodal}. Furthermore, delta and theta bands of power spectral density (PSD) of EEG when extracted from three central channels, contain information that is associated with valence and arousal \cite{frantzidis2010toward}.
In this paper, we focused on the arousal and valence scale.

\begin{figure*}[tb]
\centering
\includegraphics[width=\textwidth]{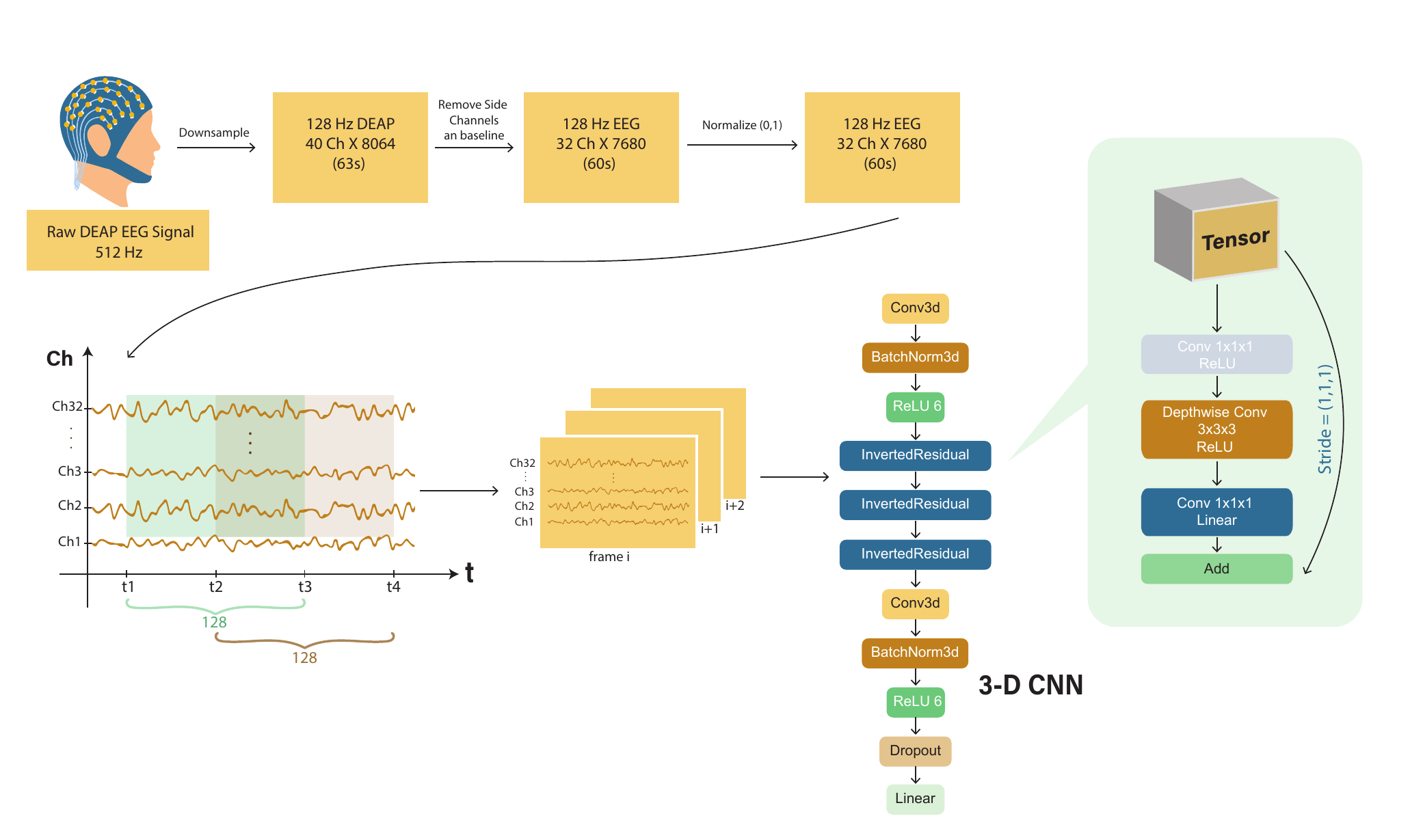}
\caption{The Data Process Steps and Proposed EEGNet Architecture}
\label{fig 1}
\end{figure*}

\section{Proposed Method}

\subsection{Efficient 3-D CNN Models with Inverted Residual Block}
The 3-D convolutional neural network expands on the traditional 2-D CNNs by adding an additional dimension. 3-D CNN is written as follows:
\begin{equation}
y^l_{i,j,k} = \sum^m_{a=0}\sum^n_{b=0}\sum^p_{c=0}\omega_{a,b,c}*x^{l-1}_{i-a,j-b,k-c}
\end{equation} Where $x^{l-1}$ is the output from the previous layer after activation applied, $\omega$ is the 3-D convolutional kernel with size $m * n * p$, and $y^l$ is the convolution output.
To the best of our knowledge, there are only a few researches that have been conducted with the use of 3-D CNNs, particularly for EEG signals. To test the benefits of using 3-D CNNs, we modified the most widely used architectures in literature, ResNet-18 \cite{he2016deep} and MobileNetV2 \cite{sandler2018mobilenetv2}, by replacing 2-D CNNs with 3-D CNNs. The use of these models on our EEG dataset resulted in excellent accuracy at the expense of very high storage and computational costs as shown in the section V. However, deploying these models on edge BCI devices is not a viable option.
\begin{table}[htbp]
\caption{Proposed models with their corresponding parameter setting}
\begin{center}
\renewcommand{\arraystretch}{1.3}
\begin{tabular}{ c c c c c}

\rowcolor{gold5}

\mc{0}{\textcolor{white}{\textbf{Model}}}& {\textcolor{white}{\textbf{\textit{Total Param.}}}}& {\textcolor{white}{\textbf{\textit{Width Factor}}}}& {\textcolor{white}{\textbf{\textit{$t$}}}}& {\textcolor{white}{\textbf{\textit{Output Neurons}}}}\\

\rowcolor{gold1}
\mc{1}{EEGNet V1}&   6.4K&	0.4&	    2&	   320\\

\rowcolor{golds}
\mc{1}{EEGNet V2}&  14.6K&	    0.5&	    3&	    640\\

\rowcolor{golds1}
\mc{1}{EEGNet V3}&  24.8K&	    0.8&	    4&	    640\\

\end{tabular}
\end{center}
\label{Table 2}
\end{table}

We created three extremely efficient network models named EEGNet V1, V2, and V3, in which we modified the block structure introduced by \cite{sandler2018mobilenetv2} by replacing the inverted residual blocks and depthwise separable convolutions to 3-D operations instead of 2-D as shown in Fig. \ref{fig 1}.  
Depthwise separable convolution decoupled the standard convolution into a $3\times3$ depthwise convolution and a $1\times1$ pointwise convolution that reduce the learning parameters and computational costs of the networks \cite{howard2017mobilenets}.
In a standard residual block, inputs are followed by multiple bottleneck layers, which are then followed by expansions. As shown with 3-D inverted residual blocks in Fig. \ref{fig 1}. We reversed the bottleneck and expansion layers, and then applied shortcut connections directly between the two bottlenecks when both the input and output tensor of the block have the same shape. The blocks also have their 3-D batch normalization layer and ReLU activation layer inside. Our models consisted of three inverted residual blocks and then were followed by another point-wise convolution layer to expand the feature maps for classification. The last two layers are a dropout layer which is included for regularization to prevent an overfitting and a fully connected linear layer for generating classification logits in which then go through a softmax function to produce final classification probabilities with two classes (positive and negative).

As shown in Table \ref{Table 2}, we tuned three hyperparameters to create three versions of EEGNet models: the expansion factor $t$, the width multiplier factor (WF), and the number of output neurons of the last convolution layer. The Expansion factor $t$ determines the number of times to expand the channels of the input tensor to the hidden expansion layers of the second and third inverted residual blocks. WF determines the overall model channel width. The number of output neurons controls the number of output neurons of the last point-wise convolution layer. The V1, V2, and V3 are in ascending order of model complexity, but increasing test accuracy as well. They are targeting edge platforms with different resource constraints and latency requirements.

\begin{figure*}[tb]
\centering
\includegraphics[width=\textwidth]{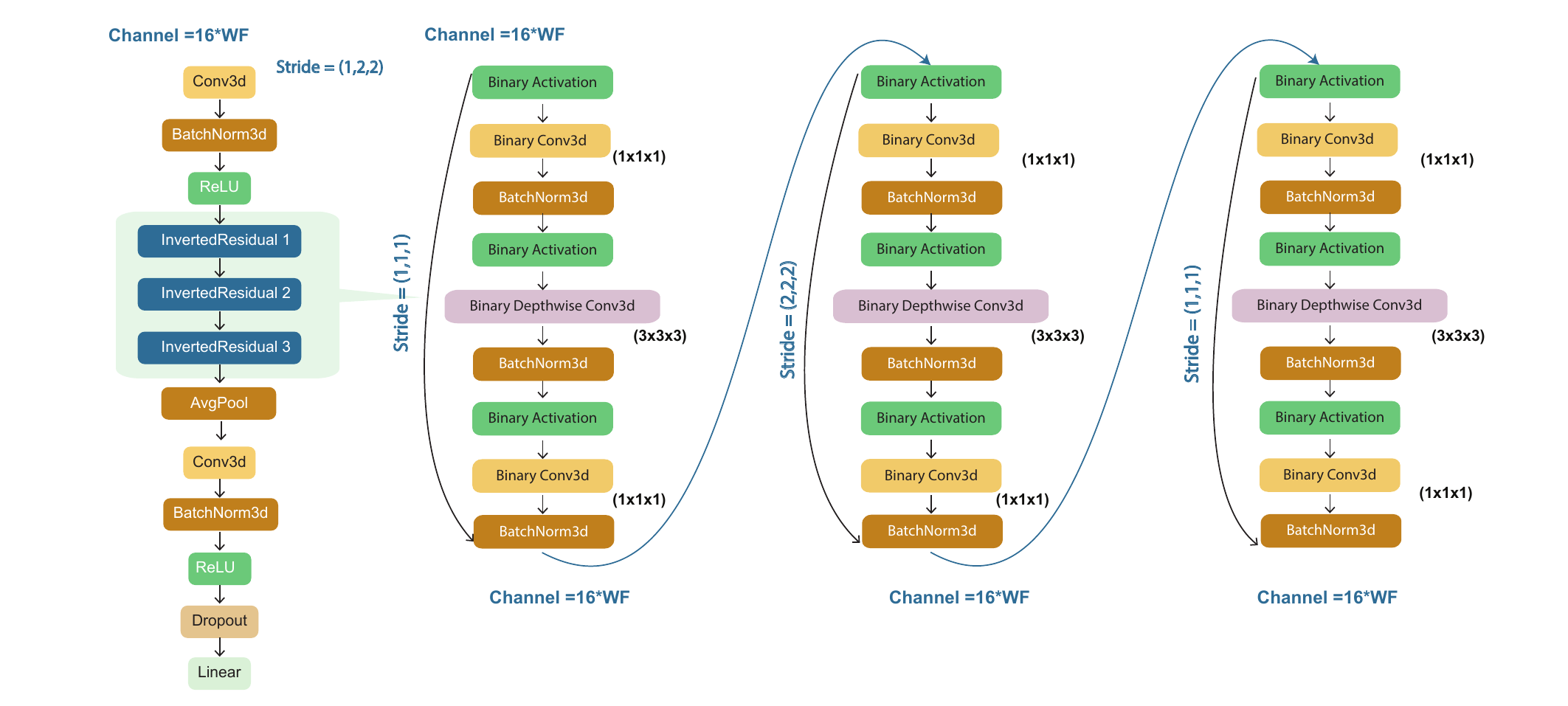}
\caption{Proposed Binary EEGNet Architecture}
\label{fig 2}
\end{figure*}
\subsection{Model Binarization}
To further compress the models, we binarized the weights and activation within the inverted residual blocks as the second stage of our framework. Binarization is an extreme case of quantization in which only one bit is used to represent the number in the weights and activation, therefore greatly reducing the computation and memory footprint. The XNOR-Net is a common binarization method that achieves up to 58x of inference speed and 32x of memory saves \cite{rastegari2016xnor}. The issue with binary neural networks is that model accuracy often degrades significantly. Hence, we introduced three techniques to improve the accuracy of the binarized models with little to no additional cost in terms of storage and computational resources.

\subsubsection{\textbf{Add real value residual connections to each block}}

Residual connections provide the possibility of constructing very deep models without performance degradation. Normally feature maps produced by convolution operation are directly calculated, but \cite{he2016deep} suggested that the convolution layers have difficulty learning the identity mapping assuming the solution is already optimal from the previous layer. With residual connection between layers, the feature map $H(x)$ now constructed by two-part: $H(x) = x + F(x)$. Where $x$ is identity mapping and $F(x)$ is the residual that is learned by the current layer, so the performance of a deep model will at worst be equal to its shallow counterpart. Our baseline models applied residual connection only when the given input channel is equal to output channel and stride is (1,1,1) as shown in Fig. \ref{fig 1}, i.e. only the first block. 

The previous residual setting is sufficient for real value baseline models, since our network architecture is not very deep. However, the binarized models will suffer significantly due to information loss when applying nonlinear activation even with a shallow network structure. Hence, for the binarized network, we preserved the real value feature map from the previous layer, and add it to the output activation. Then applied the real value residual connections to all the three blocks in the proposed models. If the convolution stride is not (1,1,1) or channels of input/output are different, we first downsampled the real value feature map to its desired shape and then added as shown in Fig. \ref{fig 2}. Those dense real value connections increased the representation capability significantly due to the limited knowledge that binarized activation maps contained. The only additional cost of adding more real value shortcuts is a small amount of element-wise addition operations and no extra memory cost because the addition operations are computed on the fly during the inference stage.

\subsubsection{\textbf{Apply channel-wise scaling factor}}
Binarize a neural network by applying the sign function $X_b = sign(X_r)$ to both the weights and the activation, resulting in crucial information loss. To address the issue, Rastegari et al. \cite{rastegari2016xnor} suggested the scaling factors $\alpha$ and $K$ to approximate floating-point weights and activation after binarization, as follows:


\begin{equation}
A*W \approx (sign(A) \odot sign(W)) \alpha K 
\end{equation}
Where $A$ and $W$ are weights and activation, $*$ is the real value convolution operator, $\odot$ represent binary convolution with XNOR and bits shift operations, $\alpha$ is a weight scaling factor such that $W_r \approx \alpha W_b$, and $K$ is the scalar factor matrix of the corresponding activation. $K$ was removed in our case due to its high computational cost and negligible impact on performance, as suggested by \cite{rastegari2016xnor}. However, We found that analytically calculated $\alpha = \frac{||W||}{n}, n=c\times w\times h$ is not optimal. It averaged out all of the weighting channels but ignored their significance and overall magnitude levels. As a result, we proposed a channel-wise scaling factor, and added the third dimension to empathize model representation ability with $\alpha_i = \frac{||W_i||}{n},\space n=k\times w\times h,\space \alpha \in \mathbb{R}^c$, where $c$ is weight channels and $k\times w\times h$ is data dimensions. The usage of channel-wise scaling factor improved the performance of binarized models significantly, as section V demonstrated.

\subsubsection{\textbf{Architecture modification for additional compression}}
In the proposed models, we binarized the layers inside the inverted residual blocks for better performance. The first and last full-precision convolution layers, which require a high computation and storage cost, become the most expensive parts of the model during training and inference. Therefore, to further compress the binarized models, we reduced the number of filters in the first channel from 32$\times$WF to 16$\times$WF. This avoided performing millions of multiplication and accumulation operations with negligible performance loss due to the redundancy nature of CNNs. Furthermore, the final full-precision convolution layer, which usually requires a high computation and storage due to its property of generating a large number of feature maps for final prediction, has been modified by adding the average pooling layer ahead. This resulted in the computations being done at $1\times 1\times 1$ spatial-temporal resolution in the final layer rather than $3\times 8\times32$. Additionally, the last stage tuning not only reduced the computation but increased the accuracy as well because the early average pooling helped to avoid location sensitivity of input features in the activation.

Combining these three methods vastly improved the performance of binarized models.

\section{Experiment and Results Analysis}
\subsection{DEAP Dataset}
DEAP \cite{koelstra2011deap} is a well-known public EEG dataset for the analysis of human affective states. It consists of 32 participants with recordings as each watched 40 one-minute-long excerpts of music videos. Then participants rated each video to the levels of valence, arousal, liking, and dominance \cite{koelstra2011deap}. Resulted in 63-second multi-channel EEG signals with 8064 sample points per channel per trial. The first 3-second of each trial is the baseline prior to the actual experiment and will be removed in our data process step. There are 40 channels that have been recorded per trial with 32 EEG channels and 8 side channels. The EEG channels are recorded using a standard international 10-20 system with 32 active AgCl electrodes \cite{koelstra2011deap}. The rated score of each video clip ranges from 1 to 9 per label.
\begin{table}[htbp]
\caption{Arousal test accuracy vs. number of frame per chunk with MobileNetV2-3D on DEAP dataset}
\begin{center}
\renewcommand{\arraystretch}{1.20}
\begin{tabular}{c| c| c| c| c}

\rowcolor{green3}
  \mc{1}{} & \mc{1}{\textbf{\textit{4}}}& \mc{1}{\textbf{\textit{6}}}& \mc{1}{\textbf{\textit{8}}}& \mc{1}{\textbf{\textit{10}}}\\ 

\rowcolor{green1}
\mc{1}{Arousal}&\mc{1}{64.2\%} & \mc{1}{99.7\%} &\mc{1}{98.5\%}&\mc{1}{97.1\%} \\

\end{tabular}
\end{center}
\label{Table 1}
\end{table}

\subsection{Data Preprocess and 3D Representation}
We first adopted a preprocessing method that is widely used in the literature \cite{hazarika1997classification}, in which we downsampled the data from 512 Hz to 128 Hz and removed Electrooculography (EOG) artifacts. Then, to eliminate unwanted frequency components, we introduced a bandpass filter, which only preserved the 4-45Hz frequency range covering Theta, Alpha, and Beta waves. The 3 seconds pre-trial baseline and the eight side channels are then removed as well. We then normalized the data for each channel of each trial to be between 0 and 1. Lastly, we divided each trial into 32 1-second data frames with a window size of 128 points and an overlapping ratio of 50\% as shown in Fig. \ref{fig 1}. In a short period of time, this method preserves the temporal stationery of EEG signals.


To prepare the datasets for 3D-CNNs, we stacked up multiple 32-channel by 1-second-long consecutive data frames to form 3-D data chunks \cite{salama2018eeg}. After experimenting with various frame sizes, we selected six frames to continue as they yielded the best accuracy, as shown in Table \ref{Table 1}. Then, we assigned each chunk the labels the same as the ground-truth labels of its corresponding trial.
In total, we constructed 25600 chunks of data with a size of 6*32*128 as the method presented. We experimented with valence and arousal perception in this paper with their corresponding labels. A threshold value of 5 was applied to assign positive and negative labels from the provided 1 to 9 scores, as this is common in the area of research literature. Finally, our proposed method is shown in Fig. \ref{fig 1}.

\subsection{Training Setting}
All models were trained on an NVIDIA Tesla V100 GPU and implemented with the PyTorch framework \cite{paszke2019pytorch}. We split the training and validation datasets with 80\% and 20\% respectively. The dropout layer was set to a 0.2 dropout rate. We trained the models for 100 epochs with a batch size of 256 and adopted Adam \cite{kingma2014adam} optimizer. The initial learning rate was set to 0.001 and The multi-step learning rate scheduler was applied with a milestone set to 75, and gamma equal to 0.5.
We used the cross entropy loss function across all models and applied additional label smoothing \cite{muller2019does} with $\epsilon = 0.1$ specifically to the binarized models. Label smoothing provides additional regularization by constructing soft labels as 
\begin{equation}
label_{soft} = (1-\epsilon)*label - \frac{\epsilon}{K}
\end{equation}
Where K is the number of label classes and was set to 2 in our cases. 

\begin{table*}[!t]
\caption{Performance (test accuracy) comparison with previous studies (DEAP)}
\begin{center}
\renewcommand{\arraystretch}{1.7}
\setlength{\tabcolsep}{16pt}
\begin{tabular}{c c c c c}

\rowcolor{green5}
\mc{1}{\textcolor{white}{\textbf{Model}}} & {\textcolor{white}{\textbf{\textit{Valence (\%)}}}}& 
{\textcolor{white}{\textbf{\textit{Arousal (\%)}}}}& {\textcolor{white}{\textbf{\textit{Parameters}}}}& {\textcolor{white}{\textbf{\textit{Memory Usage}}}}\\

\rowcolor{green6}
\mc{1}{Samara et al. \cite{samara2016feature}} & 66.90 & 66.69 & -&  -\\

\rowcolor{green7}
\mc{1}{Chao et al. \cite{chao2018recognition}} & 76.83& 75.92&	- & -\\

\rowcolor{green8}
\mc{1}{Wang et al. \cite{wang2018emotionet}}&  72.10&	73.30&	-& -\\

\rowcolor{green9}
\mc{1}{Yanagimoto et al. \cite{yanagimoto2016recognition}}&  81.16&	-&	-& -\\

\rowcolor{greens}
\mc{1}{Salama et al. \cite{salama2018eeg}}&  87.44&	88.49& 2.5M &  75.13 Mbits\\

\rowcolor{greens}
\mc{1}{ Yang et al. \cite{yang2019multi}}&  90.01&	90.65& 314K & 9.58 Mbits\\

\rowcolor{greens}
\mc{1}{ Zhao et al. \cite{zhao20203d}}&  96.43&	96.61& 170M & 5435 Mbits\\

\rowcolor{greens}
\mc{1}{Resnet18-3D}&  92.77&	97.53& 33.2M & 1012 Mbits\\

\rowcolor{greens}
\mc{1}{MobileNetV2-3D}&  99.68&	99.70& 2.4M & 71.87 Mbits\\

\rowcolor{gold1}
\mc{1}{\textbf{EEGNet V1}} &  \textbf{88.44}&	\textbf{90.04}& \textbf{6.4K} & \textbf{0.20 Mbits} \\ 
\rowcolor{gold2}
\mc{1}{\textbf{EEGNet V2}} &  \textbf{96.37}&	\textbf{96.60}& \textbf{14.6K} &  \textbf{0.45 Mbits} \\ 
\rowcolor{gold4}
\mc{1}{\textbf{EEGNet V3}} &  \textbf{99.45}&	\textbf{99.51}& \textbf{24.8K}&  \textbf{0.76 Mbits} \\ 
\rowcolor{gold1}
\mc{1}{\textbf{Bi-EEGNet V1}} &  \textbf{80.00}&	\textbf{80.95}& \textbf{3.5K+9.1K$^*$} & \textbf{0.11 Mbits} \\ 
\rowcolor{gold2}
\mc{1}{\textbf{Bi-EEGNet V2}} &  \textbf{94.42}&	\textbf{95.43}& \textbf{8.1K+33K$^*$} &  \textbf{0.28 Mbits} \\ 
\rowcolor{gold4}
\mc{1}{\textbf{Bi-EEGNet V3}} &  \textbf{99.32}&	\textbf{99.43}& \textbf{11K+130K$^*$}&  \textbf{0.46 Mbits} \\ 

\multicolumn{5}{l}{
  \begin{minipage}{5cm}
    \small Note: * denote 1-bit binary parameters.
  \end{minipage}
}\\
\end{tabular}
\end{center}
\label{Table 4}
\end{table*}

\begin{table}[htbp]
\caption{Results of 3 proposed methods for optimizing binary neural network with EEGNet V2 setting (DEAP)}
\begin{center}
\renewcommand{\arraystretch}{1.5}
\setlength{\tabcolsep}{10.5pt}
\begin{tabular}{ c c c c c}

\rowcolor{green5}

\mc{0}{\textcolor{white}{\textbf{Method}}}&{\textcolor{white}{\textbf{\textit{Arousal (\%)}}}}& {\textcolor{white}{\textbf{\textit{Improvement $\Delta$}}}}\\

\rowcolor{green9}
\mc{1}{Plain BNN}&     75.14&	-\\

\rowcolor{green9}
\mc{1}{1- Connection Real Values}&  80.30&	    +5.16\\

\rowcolor{green9}
\mc{1}{2- Channel-Wise}&  91.93&	    +16.79\\

\rowcolor{green9}
\mc{1}{3 - Tuning Last Stage}&  77.28&	    +2.14\\

\rowcolor{green9}
\mc{1}{1 \& 2}&  92.97&	    +17.83\\

\rowcolor{green9}
\mc{1}{1 \& 3}&  78.89&	    +3.75\\

\rowcolor{green9}
\mc{1}{2 \& 3}&  95.33&	    +20.19\\

\rowcolor{green9}
\mc{1}{1 \& 2 \& 3}&  95.43&	    +20.29\\

\rowcolor{gold4}
\mc{1}{Full Percision}&  96.60&	    -\\

\end{tabular}
\end{center}
\label{Table 3}
\end{table}
\subsection{Results of Baseline Models}
Our proposed method significantly outperformed previous studies without special feature extraction or signal transformation while still maintaining a compact size. As Table \ref{Table 4} shows, our EEGNet V1 achieves better results compared to  \cite{yang2019multi} and \cite{salama2018eeg} but with over 10x fewer learning parameters, which is only 6.4K. Our proposed EEGNet V2 achieves comparable performance with the state-of-the-art 3-D CNN approach \cite{zhao20203d} and 3-D ResNet18 in the experiment but with only 14.6K learning parameters, which is less than 1\% of the parameter counts compared to \cite{zhao20203d}. Ultimately, our largest V3 variant achieved an average classification accuracy of 99.5\%  with only 24.8K parameters and significantly outperformed the previous studies. To the best of our knowledge, these three variants of our proposed method outperformed the state-of-art models regarding both accuracy and model size. To elaborate more details performance analysis, We included validation precision, recall, and F1 score of each model to show its effectiveness and robustness in Table \ref{prf1}.

The 3-D CNNs with batch normalization, dense prediction, inverted residual blocks, and depthwise separable convolution made our models efficient and yields outstanding results in terms of accuracy and model size. Batch normalization helps to solve the covariance shift problem \cite{jirayucharoensak2014eeg}, and dense prediction solved the unreliability issue \cite{gunes2013categorical}. Inverted residual blocks maintained the manifolds of interest in neural networks and reduced the information loss when doing non-linear transformation \cite{sandler2018mobilenetv2}. Finally, applying depthwise separable convolution reduced the parameters and the computation complexity of the models. These advantages make efficient model deployment become possible for low power and resource-constrained device on edge.

\begin{table}[htbp]
\caption{The precision, recall and F1-score of proposed EEGnet and binary EEGNet models (DEAP)}
\renewcommand{\arraystretch}{1.76}
\setlength{\tabcolsep}{3.5pt}
\begin{center}
\begin{tabular}{ccccccc}

\rowcolor[HTML]{3CB76D} 
\cellcolor[HTML]{3CB76D}{\color[HTML]{FFFFFF} }                                  & \multicolumn{2}{c}{\cellcolor[HTML]{3CB76D}{\color[HTML]{FFFFFF} \textit{\textbf{Precision (\%)}}}} & \multicolumn{2}{c}{\cellcolor[HTML]{3CB76D}{\color[HTML]{FFFFFF} \textit{\textbf{Recall (\%)}}}} & \multicolumn{2}{c}{\cellcolor[HTML]{3CB76D}{\color[HTML]{FFFFFF} \textit{\textbf{F1 Score (\%)}}}} \\ \cline{2-7} 
\rowcolor[HTML]{E3C69C} 
\multirow{-2}{*}{\cellcolor[HTML]{3CB76D}{\color[HTML]{FFFFFF} \textbf{Models}}} & \textit{Valence}                               & \textit{Arousal}                              & \textit{Valence}                             & \textit{Arousal}                             & \textit{Valence}                              & \textit{Arousal}                              \\ \cline{1-1}
\rowcolor[HTML]{E4F3EA} 
{\color[HTML]{000000} EEGNet V1}                                                 & {\color[HTML]{000000} 90.00}                   & {\color[HTML]{000000} 90.22}                  & {\color[HTML]{000000} 89.37}                 & {\color[HTML]{000000} 92.89}                 & {\color[HTML]{000000} 89.70}                  & {\color[HTML]{000000} 91.65}                  \\
\rowcolor[HTML]{E4F3EA} 
{\color[HTML]{000000} EEGNet V2}                                                 & {\color[HTML]{000000} 96.73}                   & {\color[HTML]{000000} 96.63}                  & {\color[HTML]{000000} 96.80}                 & {\color[HTML]{000000} 97.54}                 & {\color[HTML]{000000} 96.77}                  & {\color[HTML]{000000} 97.10}                  \\
\rowcolor[HTML]{E4F3EA} 
{\color[HTML]{000000} EEGNet V3}                                                 & {\color[HTML]{000000} 99.72}                   & {\color[HTML]{000000} 99.59}                  & {\color[HTML]{000000} 99.30}                 & {\color[HTML]{000000} 99.56}                 & {\color[HTML]{000000} 99.50}                  & {\color[HTML]{000000} 99.58}                  \\
\rowcolor[HTML]{E4F3EA} 
{\color[HTML]{000000} Bi-EEGNet V1}                                              & {\color[HTML]{000000} 80.20}                   & {\color[HTML]{000000} 81.55}                  & {\color[HTML]{000000} 84.96}                 & {\color[HTML]{000000} 86.80}                 & {\color[HTML]{000000} 82.63}                  & {\color[HTML]{000000} 84.25}                  \\
\rowcolor[HTML]{E4F3EA} 
{\color[HTML]{000000} Bi-EEGNet V2}                                              & {\color[HTML]{000000} 95.02}                   & {\color[HTML]{000000} 96.90}                  & {\color[HTML]{000000} 96.46}                 & {\color[HTML]{000000} 94.98}                 & {\color[HTML]{000000} 95.75}                  & {\color[HTML]{000000} 95.88}                  \\
\rowcolor[HTML]{E4F3EA} 
{\color[HTML]{000000} Bi-EEGNet V3}                                              & {\color[HTML]{000000} 99.23}                   & {\color[HTML]{000000} 99.20}                  & {\color[HTML]{000000} 99.54}                 & {\color[HTML]{000000} 99.83}                 & {\color[HTML]{000000} 99.40}                  & {\color[HTML]{000000} 99.53}                  \\
\end{tabular}
\end{center}
\label{prf1}
\end{table}

\subsection{Results of Binarized Models}
Our model binarization algorithm adopted the piece-wise polynomial function proposed by \cite{liu2018bi} to estimate the derivative of the sign function to successfully propagation the binary models. Table \ref{Table 3} shows the effectiveness of our three proposed techniques, and results in model performance improvements of 5\%, 17\%, and 2\% respectively on arousal label with EEGNet V2 setting. The valence label showed a similar performance gain in our experiments. Combining these we achieved over 20\% improvement over the plain binary model while still maintaining similar storage and computational cost. As shown in Table \ref{Table 4}, the binarized models with v2 and v3 settings achieved comparable performance to their full precision counterparts while further compressing the size over 40\%. The memory usage (i.e. model size) was calculated by summing up the 32-bit multiples the number of real-valued parameters and 1-bit multiples the number of binary parameters in each model.
Hence, our binarized models can take advantage of the proposed method and speed up significantly when deploy to potential edge systems while still maintaining state-of-the-art performance.     


\section{Conclusion}
In this paper, we proposed a two-stage 3-D CNN framework that specialized in emotion recognition tasks using time-domain EEG signals.
The framework extracted spatiotemporal feature representations automatically. The public DEAP dataset was used to conduct experiments with data processing techniques to form the 3-D inputs. Our models showed superior performance on classifying both valence and arousal perceptions, which can be easily processed to actual human emotions afterward. The introduced baseline EEGNet V1, V2, and V3 in the first stage achieved average classification accuracies of 90\%, 96.6\%, and 99.5\% respectively with a small number of learning parameters and compact model size. In the second stage, we binarized these models with three novel techniques to perform further compression and help to take advantage of efficient bitwise operation. Model binarization saved over 40\% of storage costs and computational resources compared to the baseline EEGnet, while still keeping comparable performance to their full precision counterparts. This paper mainly focused on the algorithmic and efficient model design. In future work, we will provide more comprehensive experiments on real edge environments with detailed latency and energy consumption analysis. A neural architecture search (NAS) strategy based on Multi-Objective Bayesian Optimization (MOBO) will also be introduced to further improve the existing model design to balance various device constraints and needs.
Finally, the efficient models we presented helped the real-time deployment of a precise human emotion recognition system in a resource-constrained environment become a viable option.


\bibliography{refernces.bib}{}

\bibliographystyle{IEEEtran}

\end{document}